\documentclass[12pt]{article}

\begin{document}
\title{ Action for particles faster than  light\\}

\author{ Juan M. Romero\thanks{jromero@correo.cua.uam.mx},
Jos\'e A. Santiago \thanks{jsantiago@correo.cua.uam.mx},
O. Gonz\'alez-Gaxiola\thanks{ogonzalez@correo.cua.uam.mx},
\\[0.5cm]
\it Departamento de Matem\'aticas Aplicadas y Sistemas,\\
\it Universidad Aut\'onoma Metropolitana-Cuajimalpa\\
\it M\'exico, D.F  01120, M\'exico\\[0.3cm]} 

\pagestyle{plain}

\maketitle

\begin{abstract}
A general action for  particles faster than light is presented.   It is demonstrated that this action is invariant under reparametrizations. For several cases,  it is  shown that in the high velocity  regime the  action is invariant under anisotropic space-time transformation and  at quantum level the system has  fractal behavior.  For those cases,  it is  shown  that the action describes a particle in Finsler geometry and  equivalent to  one dimensional field theory in a curved space,  where the metric depends on temporal derivatives.

\end{abstract}

\section{Introduction}
Special relativity has been one of most important theories in physics. However, 
recently OPERA collaboration has reported particles  faster than light  \cite{opera:gnus}.  Some authors think that we have to wait another experiment to 
confirm that result while others think that  such experiment  might has mistakes or  that has not been  well interpreted  \cite{glashow:gnus}.  
But, if that experiment is right,  special relativity  must  be changed.  There are  several  theories with breaking Lorentz symmetry. 
For example, P. Ho\v{r}ava has proposed a gravity theory with  breaking Lorentz symmetry, which   is  renormalizable  \cite{horava:gnus}. 
In High Energy Physics there are some extensions of the  standard model  that consider 
Lorentz symmetry violation \cite{kostelecky:gnus,anselmi:gnus}.  
Also, it is worth to mention that MOND theory  is an alternative propose to dark matter \cite{milgrom:gnus,bekenstein1:gnus} and   breaks Lorentz symmetry too \cite{bekenstein2:gnus}. 
Many of these theories  have important properties, for example, Ho\v{r}ava' s gravity has an alternative mechanism to inflation   \cite{brandenberger:gnus}  and can explain some cosmological phenomena without dark matter \cite{mukohyama:gnus}. Doubly Special Relativity (DSR) is another interesting theory with  breaking Lorentz symmetry \cite{amelino1:gnus,amelino2:gnus,amelino3:gnus}. Other works with breaking Lorentz symmetry can be seen in \cite{bola:gnus,bola1:gnus,bola2:gnus,bola3:gnus,bola4:gnus,bola5:gnus,bola5a:gnus,bola6:gnus,bola7:gnus,bola8:gnus,bola9:gnus,bola10:gnus,bola11:gnus,bola12:gnus,bola13:gnus,bola14:gnus,bola15:gnus,bola16:gnus,saridakis:gnus}.\\

According to special relativity, the action for a particle with mass $m$ and velocity $u$ is
\begin{eqnarray}
S=-m\int dt \sqrt{1-u^{2}},\label{eq:ac0}
\end{eqnarray}
where $t$ is the time. In this paper we take light velocity $c=1.$ \\ 

If $u$ is near from $1,$ we can take $u=1+\delta,$ where $|\delta|<<1.$ In this case we find
$\sqrt{1-u^{2}}\approx \sqrt{-2\delta}.$  We can see that if $\delta >0,$ the particle is  faster than light and the action $S$ is nonsense, for example  OPERA  reported 
$\delta=\left (2.37 \pm 0.32  (stat.)_{-0.24}^{+0.34} (sys.)\right) \times 10^{
-5} .$ 
 However,  if there are particles faster than light,  this action must be only an approximation of  
another one that  makes sense if $u> 1$. In that case we have to change
\begin{eqnarray}
\sqrt{1-u^{2}}
\end{eqnarray}
for another term. Notice that in this case, Lorentz transformation is only a limit of another symmetry.\\

In this work we present a general action for  particles faster than light and  demonstrate  that it is invariant under reparametrizations. First we propose a simple model and show that in high velocity regime its  action is invariant under anisotropic space-time transformation and  at quantum level the system has  fractal behavior. Then, it is possible that in the high energy regimen a particle has that kind of behavior. Interestingly,  Ho\v{r}ava gravity has fractal properties   \cite{horava2:gnus}  and  other authors
argued that at Plank scale the space-time has fractal properties as well \cite{fractal1:gnus,fractal2:gnus,fractal3:gnus}. Also, we show that these  kind of actions
are equivalent to  one dimensional field theory in a curved space,  where the metric depends on temporal derivatives. Moreover we show that there is a relation between those actions
and Finsler geometry. It is worth mentioning that recently Finsler geometry was proposed  as an generalized Minkowski  geometry \cite{girelli1:gnus,girelli2:gnus,girelli3:gnus}. \\

This work is organized in the following way:  In section $2$ the general action is presented; in section $3$ a simple model is studied;
in section $4$ the relation between  our model and Finsler geometry  is analyzed  and finally in section $5$ a brief summary is given.

\section{Action}

In this section we  propose a new action that  describes  particles faster than light and  reduces to the usual relativistic one.\\

The action for a  relativistic  particle is 
\begin{eqnarray}
S=-m \int d \tau  \sqrt{ \left( \frac{dt}{d\tau} \right)^{2}- \left(\frac{d\vec  x}{d\tau} \right)^{2} }, \quad \dot t=\frac{dt}{d\tau},
\label{eq:act1}
\end{eqnarray}
which is invariant under repametrizations   in $\tau$: 
\begin{eqnarray}
\tau \to \tau=\tau(\omega). 
\end{eqnarray}
Notice that if $\tau =t$ we obtain  (\ref{eq:ac0}).  \\

Let $f$ be a function that if $u\leq 1,$ then  $f(u^{2})  \approx   \sqrt {1-u^{2}}.$ But if $u>1,$ it makes sense. Thus, the action  
\begin{eqnarray}
S=-m \int d \tau L, \qquad L=\frac{dt}{d\tau} f(u^{2}), \quad u^{2} =\left( \frac{d\vec x}{d\tau} \frac{1}{  \frac{dt}{d\tau}} \right )^{2}
\label{eq:act2}
\end{eqnarray}
is a generalization of (\ref{eq:act1}). We can see  that (\ref{eq:act2})  is also invariant under reparametrizations  in $\tau.$
Now, using (\ref{eq:act2}) we find
\begin{eqnarray}
E&=&-P_{t}=-\frac{\partial L}{\partial \dot t}= m \left(f(u^{2})-2u^{2} \frac{df(u^{2})}{du^{2}}\right),\label{eq:ener}\\
P_{i}&=&  \frac{\partial L}{\partial \dot x^{i} }=-2m  \frac{df(u^{2})}{du^{2}} \frac{\dot x_{i}}{\dot t},\qquad P=2mu\left|\frac{df(u^{2})}{du^{2}}\right|,\label{eq:momen}
\end{eqnarray}
then
\begin{eqnarray}
H=P_{t}\dot t +\vec P\cdot\frac{d\vec x}{d\tau} -L=0,
\end{eqnarray}
and 
\begin{eqnarray}
\frac{E}{P}&=&\left[ \frac{ f(u^{2})}{ 2u \left|\frac{df(u^{2})}{du^{2}}\right|}-\left(\frac{ \frac{df(u^{2})}{du^{2}}}   {\left |\frac{df(u^{2})}{du^{2}}\right|} \right) u\right].
\end{eqnarray}
When this equation  is invertible, we can express $u$ as a function of $E$ and $P,$ namely  $u=u(E,P).$
Then, using (\ref{eq:ener}) and (\ref{eq:momen}) we have a constraint 
\begin{eqnarray}
\phi(E,P)=0.
\end{eqnarray}
Therefore,  using the Dirac's  method \cite{dirac:gnus},  the extended Hamiltonian is given by
\begin{eqnarray}
H_{ext}= \lambda \phi(E,P),
\end{eqnarray}
where $\lambda$ is a Lagrange multiplier. Then, the Hamiltonian action is 
\begin{eqnarray}
S_{H}=\int d\tau \left( P_{t}\dot t+ P_{i}\dot x_{i} - \lambda \phi\left(P_{t},P_{i} \right)\right).
\end{eqnarray}

\subsection{Alternative actions}
The action (\ref{eq:act2}) has at least  three alternative actions. First,  we  consider 
\begin{eqnarray}
S_{I}=-\frac{1}{2} \int d\tau \dot t \left[ \frac{\left(f(u^{2})\right)^{2} }{\lambda }+\lambda m^{2}\right] .
\label{eq:act3}
\end{eqnarray}
The  equation of motion  for $\lambda$ implies  
\begin{eqnarray}
\lambda =\frac{ f(u^{2})}{m},
\end{eqnarray}
substituting  this result in (\ref{eq:act3}) we obtain (\ref{eq:act2}).  We can see that, unlike (\ref{eq:act3}),  in this action the case $m=0$ makes sense.  \\

Another alternative action is 
\begin{eqnarray}
S_{II}=-m\int d\tau \dot t \left[ f(\lambda) +\left(u^{2} -\lambda\right) \frac{df(\lambda)}{d\lambda } \right].
\label{eq:act4}
\end{eqnarray}
For this, the  equation of motion   for $\lambda$ gives  
\begin{eqnarray}
\lambda =u^{2},
\end{eqnarray}
substituting  this result in (\ref{eq:act4}) we obtain (\ref{eq:act2}).  \\

Now, let us consider 
\begin{eqnarray}
S_{III}=-\frac{1}{2} \int d\tau \dot t \left[ \frac{\left(f(\beta)\right)^{2} +(u^{2}-\beta)\frac{d\left(f(\beta)\right)^{2} }{d\beta} }{\lambda }+
\lambda m^{2}\right] .\label{eq:act5}
\end{eqnarray}
Using the equations of motion for $\lambda$ and $\beta,$ we  get (\ref{eq:act2}).\\

\section{Anisotropic cases }

A simple model which describes particles faster than light is
\begin{eqnarray}
S=-m \int d \tau \frac{dt}{d\tau}  \sqrt{1-u^{2}+\alpha u^{2n}}  .
\label{eq:leo1}
\end{eqnarray}
Notice that, when $u>>1,$ we obtain
\begin{eqnarray}
S\approx -m \sqrt{\alpha} \int d \tau \frac{ \left( \dot x^{2}\right)^{\frac{n}{2} }}{\dot t^{n-1} }.
\end{eqnarray}
This action is invariant under anisotropic transformations
\begin{eqnarray}
t \quad \to\quad b^{z} t, \qquad  \vec x  \quad \to\quad b \vec x, \qquad 
z=\frac{n}{n-1}.
\end{eqnarray}
Therefore,   we get 
\begin{eqnarray}
P_{t}&=& m\sqrt{\alpha}(n-1) \frac{ \left( \dot x^{2}\right)^{\frac{n}{2} }}{\dot t^{n} },\\
P_{i}&=& -m\sqrt{\alpha}  \frac{ \left( \dot x\right)^{n-2 }}{\dot t^{n-1} } \dot x_{i}
\end{eqnarray}
and the dispersion relation
\begin{eqnarray}
E^{2}= \frac{ (n-1)^{2} }  {  \left(m^{2} \alpha\right) ^{\frac{1}{n-1} } }\left( P^{2}\right)^{\frac{n}{n-1} } .
\end{eqnarray}
Then, in the quantum theory  the wave equation  is given by
\begin{eqnarray}
-\frac{\partial^{2} \psi }{\partial t^{2}}= \frac{ (n-1)^{2} }  {  \left(m^{2} \alpha\right) ^{\frac{1}{n-1} } }\left(- \nabla^{2} \right)^{\frac{n}{n-1} } \psi,\qquad \hbar=1 .
\end{eqnarray}
It is a fractional wave equation \cite{fractional1:gnus}. Fractional-like wave  equations describe systems with fractal properties \cite{fractional2:gnus,fractional3:gnus}. Then, it is possible that 
in the high energy regimen a particle has fractal properties. It is worth to mention that Ho\v{r}ava gravity has similar  properties   \cite{horava2:gnus}  and  other authors
argued that at Plank scale the space-time has fractal properties as well \cite{fractal1:gnus,fractal2:gnus}.

\section{Relation with Finsler geometry}  

Recently Finsler geometry was proposed as an alternative to the Minkowski one in the regime
close to the Plank scale \cite{girelli1:gnus,girelli2:gnus,girelli3:gnus,girelli4:gnus}. This geometry has been proposed as brackground for
some theories, like  MOND \cite{finsler1:gnus}, very special relativity \cite{finsler2:gnus} and DSR \cite{finsler3:gnus}. In this geometry the action is
\begin{eqnarray}
S=\int d\tau  F( x^{\mu},\dot x^{\mu}), 
\label{eq:fins-ac0}
\end{eqnarray}
where   $F( x^{\mu},\dot x^{\mu})$ is a homogeneous function of one degree of $\dot x^{\mu}.$ 
For example,  if   $\gamma_{\mu\nu}(x^{\mu},\dot x^{\mu})$ is a homogeneous function of zero degree of $\dot x^{\mu},$ then 
\begin{eqnarray}
S=\int d\tau \sqrt{\gamma_{\mu\nu}(x,\dot x)\dot x^{\mu}\dot x^{\nu}}
\label{eq:fins-ac}
\end{eqnarray}
is Finsler-like action.\\

In this section we show a relation between action (\ref{eq:leo1}) and the Finsler one (\ref{eq:fins-ac}). 
Let us notice that the action (\ref{eq:leo1}) can be written as
\begin{eqnarray}
S=-m \int d \tau  \sqrt{g_{\mu\nu}\dot x^{\mu}\dot x^{\nu}} 
\label{eq:1}
\end{eqnarray}
where 
\begin{eqnarray}
g_{\mu\nu}=\eta_{\mu\nu}+Z_{\mu\nu},\qquad Z_{\mu\nu}= \left(
\begin{array}{cccc}
 0 & \cdots\\ 
 \vdots  & \alpha \left(\frac{\dot x^{2}}{\dot t^{2}}\right)^{n-1} \delta_{ij} \\
 \end{array}
\right).
\end{eqnarray}
This metric depends on velocity and if $\dot x^{\mu}\to \Omega \dot x^{\mu},$ then 
$g_{\mu\nu}\to g_{\mu\nu}.$  Therefore  (\ref{eq:leo1}) is a Finsler-like action. \\

In general, if $h(u^{2})$ is a regular function, the action  
\begin{eqnarray}
S=-m \int d \tau  \dot t \sqrt{1-u^{2} +h(u^{2})}
\end{eqnarray}
can be expressed like
\begin{eqnarray}
S=-m \int d \tau  \sqrt{g_{\mu\nu}\dot x^{\mu}\dot x^{\nu}}
\label{eq:finsler-gen} 
\end{eqnarray}
with
\begin{eqnarray}
g_{\mu\nu}=\eta_{\mu\nu}+Z_{\mu\nu},\qquad Z_{\mu\nu}= \left(
\begin{array}{cccc}
 0 & \cdots\\ 
 \vdots  & \frac{h(u^{2})}{u^{2}} \delta_{ij} \\
 \end{array}
\right).
\end{eqnarray}
Also, this is Finsler-like action. Notice that, for several cases, the general action (\ref{eq:act2})   is Finsler-like action too.
 Then it is possible that for particles faster than light Finsler geometry replaces the Minkowski geometry. \\

Additionally, the action (\ref{eq:finsler-gen}) can be written as 
\begin{eqnarray}
S=\int d \tau \sqrt{G} \left( G^{\tau \tau} g_{\mu\nu} \partial_{\tau} x^{\mu}\partial_{\tau}x^{\nu}+\Lambda
  \right), \qquad G_{\tau\tau}=\lambda^{2},\qquad \Lambda= m^{2} 
\end{eqnarray}
This action can be  interpreted as a  1-dimensional  field  in a  curve space with metric $G_{\tau\tau},$ cosmological constant $\Lambda$ and background $g_{\mu\nu}.$\\

\section{Summary}
In this work we presented  a general action for  particles faster than light.   We demonstrated  that this action is invariant under reparametrizations. For several cases,  it was  shown that in the high velocity  regime the  action is invariant under anisotropic space-time transformation and  at quantum level the system has  fractal behavior. It is worth to mention that other authors have  proposed theories,  like Ho\v{r}ava gravity, with fractal properties. Moreover, for  the same cases,  it was  shown  that the action describes a particle in Finsler geometry and  equivalent to  one dimensional field theory in a curved space,  where the metric depends on temporal derivatives.\\

Then, if there are particles faster than light, we would ample ground for further research in spacetime-symmetry physics. However, we have to wait more experiments to find  new physics, in particular to 
know the action for a free particle.

\end{document}